\documentclass[12pt]{article}
\usepackage{amsfonts}
\usepackage{amssymb}

\newcommand{\be}{\begin{equation}}
\newcommand{\ee}{\end{equation}}
\newcommand{\bea}{\begin{eqnarray}}
\newcommand{\eea}{\end{eqnarray}}
\newcommand{\bt}{\begin{tabular}}
\newcommand{\et}{\end{tabular}}
\newcommand{\ba}{\begin{array}}
\newcommand{\ea}{\end{array}}

\newcommand{\bvec}{\mathbf}
\newcommand{\mrm}{\mathrm}
\begin{document}
\setcounter{page}{0}
\thispagestyle{empty}
\baselineskip=20pt
%---------------------------------------------------------------------------

\hfill{
\begin{tabular}{l}
DSF$-$96/25 \\
INFN$-$NA$-$IV$-$96/25
\end{tabular}}

\bigskip\bigskip

\begin{center}
\begin{huge}
{\bf Classical $v_{gr} \neq c$ solutions of Maxwell equations and the 
tunneling photon effect }
\end{huge}
\end{center}

\vspace{2cm}

\begin{center}
{\Large
Salvatore Esposito\\} 
\end{center}

\vspace{0.5truecm}

\normalsize
\begin{center}
{\it
\noindent
Dipartimento di Scienze Fisiche, Universit\`a di Napoli,
Mostra d'Oltremare Pad. 19, 80125 Napoli Italy}\\
\noindent
and\\
INFN, Sezione di Napoli,
Mostra d'Oltremare Pad. 20, I-80125 Napoli Italy\\
\noindent
e-mail: sesposito@na.infn.it
\end{center}

\vspace{3truecm}

\begin{abstract}
We propose a very simple but general method to construct solutions of 
Maxwell equations propagating with a group velocity $v_{gr} \neq c$. 
Applications to wave guides and a possible description of the known 
experimental evidences on photonic tunneling are discussed.
\end{abstract}

%\vspace{1truecm}
%\noindent
%PACS 12.10 - Unified field theories and models.\\
%PACS 12.15.Ff - Quark and lepton masses and mixing.\\
%PACS 13.15 - Neutrino interactions.\\
%PACS 95.90 - Other topics in astronomy and astrophysics.

\newpage

\section{Introduction}

\indent
It is a well known tool that solutions of Maxwell equations in 
vacuum
\bea
\nabla \cdot {\bf E} & = & 0  \;\;\;\;\;\;\;\;\;\;\;\;\;\;\;\;\;\;
\nabla \times {\bf E} \; = \; - \, \frac{1}{c} \, \frac{\partial {\bf 
B}}{\partial t} \nonumber \\
& & \label{11} \\
\nabla \cdot {\bf B} & = & 0  \;\;\;\;\;\;\;\;\;\;\;\;\;\;\;\;\;\;
\nabla \times {\bf B} \; = \;  \frac{1}{c} \, \frac{\partial {\bf 
E}}{\partial t} \nonumber 
\eea
are electromagnetic waves:
\be
\nabla^2 \phi \; - \; \frac{1}{c^2} \, \frac{\partial^2 \phi}{\partial 
t^2} \; = \; 0
\label{12}
\ee
$\phi$ being any component of {\bf E},{\bf B}. From eq. (\ref{12}) 
follows that the velocity of advancement of the related wave surface 
(front velocity) is $c$ \cite{Levi}, the speed of light in vacuum.
For a wave packet, it is of particular relevance also the concept of 
``group velocity'', being the velocity $v_{gr}$ with which the maximum of 
the wave packet propagates. To this regard, it is commonly believed 
that the group velocity of an electromagnetic wave in vacuum takes 
always the value $c$. However, there is no proof \cite{Bosanac} that 
electromagnetic waves with $v_{gr}=c$ are the only solutions of eq. 
(\ref{11}) or (\ref{12}) and, in particular, there are many works 
\cite{Batetal,RodVaz} dealing with both $v_{gr} < c$ and 
$v_{gr} > c$ solutions.

On the experimental side, many recent experiments on 
photonic tunneling conducted with different techniques and in 
different ranges of frequency 
\cite{Nimtz,diss,Stein,Spiel,Ranfagni} 
have shown 
that, in peculiar conditions, (evanescent) electromagnetic waves 
travel a barrier with (group) velocity $v_{gr} > c$. In these experiments, 
as remembered in \cite{Causality}, only the group velocity can be 
determined, so that these results obviously do not violate Einstein 
causality because, according to Sommerfeld and Brillouin 
\cite{Brillou}, it is the front velocity (not group velocity) to be 
relevant for this and, as stated above, the Maxwell theory predicts 
that electromagnetic waves in vacuum have always a constant front 
velocity equal to $c$. 

From the theoretical point of view, the difficulty in the 
interpretation of the experimental results lies mainly in the fact 
that in the barrier traversal no group velocity can be defined, the 
wave number being imaginary (evanescent waves), so that the time 
required for the traversal (directly measured) is not univocally 
defineable  
\footnote{ Incidentally, we quote the introduction of the concept of 
``phase time'' which is a generalization of that of group velocity, 
and it is applicable in this case.}\cite{Rev}.

In this paper we want to give a simple but general method for building 
up solutions of Maxwell equations propagating with group velocity 
$v_{gr} \neq c$ that generalizes previous calculations 
\cite{RodVaz} and to show how the theory could describe,
almost qualitatively, the experimental evidence for 
$v_{gr} > c$ electromagnetic propagation. To this purpose, in the following 
section the general formalism is outlined while in section 3 this 
formalism is applied to electromagnetic signals in a wave guide. In 
section 4 the experimental evidence on photonic tunneling is reviewed 
and discussed, and its possible description in our approach is 
shown together with future possible tests. 

However, it is worth noting that the theory developed in section 2 can 
be applied as well to propagation of usual electromagnetic waves in a 
medium (making the substitution $c \rightarrow c/n$, $n$ being the 
refractive index); in this case our approach can be an alternative 
one to well known phenomena such as, for example, plasma 
oscillations, Cherenkov effect and so on.

\section{General formalism}
\indent
In the experiments performed on superluminal barrier traversal, the 
wavelength $\lambda$ of the incident electromagnetic signal is always 
greater than or, at least, of the some order of a typical length of the 
experimental apparatus 
\cite{Nimtz,diss,Stein,Spiel,Ranfagni}. 
So, we have to use an approximation just opposite to 
the eikonal approximation, and we search for solution of eq.(\ref{12}) 
in the form
\be
\phi (\bvec{x},t) \; = \; \phi_0 (\bvec{x},t;\bvec{k},\omega) \, 
e^{i (\bvec{k} \cdot \bvec{x} - \omega t)}
\label{21}
\ee
with an amplitude $\phi_0$ strongly dependent on 4-position 
$\bvec{x},t$. Substituting in eq.(\ref{12}) we find the equation 
satisfied by $\phi_0$:
\be
\partial^2 \phi_0 \, + \, 2 i \left( \frac{\omega}{c^2} \frac{\partial 
\phi_0}{\partial t} \, + \, \bvec{k} \cdot \bvec{\nabla} \phi_0 
\right) \, + \, \left( \frac{\omega^2}{c^2} \, - \, k^2 \right) \phi_0 
\, = \, 0
\label{22}
\ee
For illustrative purpose we consider one-dimensional problems; 
taking $\phi_0$ dependent only on $\zeta \, = \, 
z \, - \, v t$ (and $\omega,k$), where $v$ is a given parameter (as 
well as $\omega,k$) that we will show to be identified with the group 
velocity, equation (\ref{22}) now assumes the form
\be
\left( 1 \, - \, \frac{v^2}{c^2} \right) \frac{\partial^2 
\phi_0}{\partial \zeta^2} \; + \; 2i \left( k \, - \, \frac{v 
\omega}{c^2} \right) \frac{\partial \phi_0}{\partial \zeta} \; + \; 
\left( \frac{\omega^2}{c^2} \, - \, k^2 \right) \phi_0 \; = \; 0
\label{24}
\ee
Here we observe that, if $\omega \, = \, ck$ (and then $v=c$), 
eq.(\ref{24}) is automatically satisfied, whatever being the 
dependence of $\phi_0$ on 4-position; however, for $\phi_0$ 
$\zeta$-dependent, these solutions are not the only ones. As we will 
show now, there are also $v \neq c$ solutions; these can be obtained 
imposing a ``quantification rule'' (in the sense that it quantifies 
the effective dispersion relation) specifying how strong the eikonal 
approximation is violated. 

Let us impose that $\phi_0$ in addition to eq. (\ref{24}) satisfies also
\be
\left( 1 \, - \, \frac{v^2}{c^2} \right) \frac{\partial^2 
\phi_0}{\partial \zeta^2} \; = \; - \, \Omega^2 \, \phi_0
\label{25}
\ee
with a given constant $\Omega^2 > 0$. In other words, we costrain $\phi_0$
to satisfy a harmonic motion equation whose frequency $\gamma \Omega$ 
($\gamma \, = \, 1/\sqrt{1 - v^2/c^2}$) quantifies the effective 
dispersion relation. In fact, substituting eq. (\ref{25}) in (\ref{24}) we 
find that 
\be
\phi_0 \; = \; \phi^{\prime}_0 \, e^{\pm \gamma \Omega \zeta}
\label{26bis}
\ee
is solution of eq. (\ref{25}) with
\be
\omega^2 \; = \; c^2 \, \left( k^2 \, + \, \Omega^2 \right)
\label{26}
\ee
and
\be
v \; = \; \frac{c^2 \, k}{\omega} \; = \; c \, \sqrt{1 \, - \, 
\frac{c^2 \Omega^2}{\omega^2}} \; < \; c
\label{27}
\ee
For real $v$ we see that, for fixed $\omega$, $\phi$ is a product of 
waves; in this sense, provided $\gamma \Omega v \, \ll \, \omega$ and 
noted that
\be
v \; = \; \frac{\partial \omega}{\partial k} |_{\omega} 
\;\;\; ,
\label{23}
\ee
$v$ may be interpreted as a group velocity.

Notice that for very small $\omega$ and {\it finite} $\Omega$ both $k$ 
and $v$ are imaginary, so that in this case, in a common language, we 
are describing evanescent waves.

Now, let us consider a wave packet with a given spectrum in $\omega$
\bea
\phi (z,t) & = & \int d \omega \, \phi^{\prime}_0 (\omega) \, e^{\pm i  
\, \gamma (v) \, \Omega \, (z - v(\omega) t)} \, e^{i \,
(k(\omega) z - \omega t)} \; =  \nonumber \\
& = &  \int d \omega \, \phi^{\prime}_0 (\omega) \, e^{i
(k^{\prime} z - \omega^{\prime} t)}
\eea
with $k^{\prime} (\omega) = k(\omega) \pm \gamma \Omega$, 
$\omega^{\prime} (\omega) = \omega \pm \gamma \Omega \, v(\omega)$. 
Suppose that in the spectrum $\phi^{\prime}_0 (\omega)$ of the wave 
packet one has a pronounced maximum $\omega_0$; expanding the 
integrand in $\omega$ around $\omega_0$ we get
\bea
\phi (z,t) & \approx & \int d \omega \, \phi^{\prime}_0 (\omega) \, 
e^{ i  
\, \{ [ k_0 \, \pm \, \gamma \, \Omega \, + \, \frac{\partial k}{\partial 
\omega} |_{\omega_0} (\omega \, - \omega_0)]z \, - \, [\omega_0 \, \pm 
\, \gamma \, \Omega \, \frac{\partial \omega}{\partial k}|_{\omega} 
\, + \, (\omega \, - \, \omega_0 )]t \} } \nonumber \\
& = & e^{i (k_0 z - \omega_0 t)} \, \int d \omega \, \phi^{\prime}_0 
(\omega) \, e^{i \, \{ [ \pm \gamma \, \Omega \, + \, 
\frac{\partial k}{\partial \omega} |_{\omega_0} (\omega \, - \omega_0)]z 
\, - \, [\pm \, \gamma \, \Omega \, \frac{\partial \omega}{\partial k}|_{
\omega} \, + \, (\omega \, - \omega_0)]t \} }
\eea
($k(\omega_0) \, = \, k_0$). From this, it is easy to obtain the group 
velocity of the wave packet:
\be
v_{gr} \; = \; \lim_{\omega \rightarrow \omega_0} \, 
\frac{\pm \, \gamma \, \Omega \, \frac{\partial \omega}{\partial k}|_{
\omega} \, + \, (\omega \, - \omega_0)}{\pm \gamma \, \Omega \, + \, 
\frac{\partial k}{\partial \omega} |_{\omega_0} (\omega \, - \omega_0)}
\; = \; \frac{\partial \omega}{\partial k} |_{\omega_0} \; = \; 
v(\omega_0)
\ee
Let us stress that for $\Omega \neq 0$ we have $v(\omega_0) < c$ (see 
eq. (\ref{27})); we then obtain solution 
of Maxwell equations propagating with a 
group velocity lower than $c$ depending on 
``construction'' boundary condition 
(see below) through $\Omega^2$. In fact, the space-time evolution of 
wave amplitude, and then $\Omega^2$ in eq. (\ref{25}), can be 
``constructed'' experimentally and the dispersion relation (\ref{26}) 
or the group velocity in eq. (\ref{27}) can be further measured. 

Now, instead of eq. (\ref{25}), let us impose the following 
equation
\be
\left( \frac{v^2}{c^2} \, - \, 1\right) \frac{\partial^2 
\phi_0}{\partial \zeta^2} \; = \; - \, \Omega^2 \, \phi_0
\label{28}
\ee
again with $\Omega^2 >0$. Substituting in eq. (\ref{24}) we now easily 
find that $\phi_0$ is solution of eq. (\ref{28}) with a dispersion 
relation
\be
\omega^2 \; = \; c^2 \, \left( k^2 \, - \, \Omega^2 \right)
\label{29}
\ee
propagating with a superluminal group velocity
\be
v \; = \; \frac{c^2 \, k}{\omega} \; = \; c \, \sqrt{1 \, + \, 
\frac{c^2 \Omega^2}{\omega^2}} \; > \; c
\label{210}
\ee
Note that even in this case $\phi_0$ satisfies a harmonic motion 
equation (\ref{28}); now, however, both $k$ and $v$ are always real, 
so that we have no evanescent waves. 

Incidentally, let us observe \cite{RodVaz} that our subluminal 
solutions can be equivalently constructed by requiring that $\phi_0$ 
satisfies an Helmoltz wave equation on which solutions a Lorentz boost, 
for example in the $z$ direction, is applied.
However, we again point out the fact that a quantification rule can be 
realized only by construction of a given experiment. 
%an effective dispersion 
%relation is peculiar of a given experiment.

Let us now discuss this last point and, in particular, the 
introduction of the parameter $\Omega$ in the effective dispersion 
relations. For simplicity, we confine ourselves only to subluminal
group velocities, but the same will remain valid also for the 
superluminal case.

Our method is based mainly on eq. (\ref{25}) which admits, for given 
initial conditions, an univocal solution $\phi_0$ in (\ref{26bis}) and 
then $\phi$ in (\ref{21}). This wave, product of waves, is what one 
can measure and is  univocally determined by the experimental setup 
employed. In this sense, the parameter $\Omega$ is given ``by 
construction'' once the experiment is given. For example, in 
experiments operating in the eikonal approximation regime (this is not 
the case of photonic tunneling experiments), namely when the 
wavelength $\lambda$ of the electromagnetic signal is much lower than 
a typical length of the experimental apparatus, the parameter $\Omega$ 
is zero. On the contrary, in 
experiments with wave guides one can experimentally determine 
$\Omega$ by measuring the (effective) ``cutoff'' frequency 
(as we will show in the next section, the introduction of $\Omega$ 
leads to a modification of the cutoff frequency of the wave guide).

Finally, we stress that the actual theory can be consistently tested 
only in problems in more than one dimension. In fact, let us consider, 
for example, an incident signal {\rm on} a wave guide: it is well 
known that the signal effectively propagates (with a finite group 
velocity) only in the axial direction, while across the guide a 
non-physical signal propagates with an arbitrarily high group velocity.
In the following section we will show how our approach can be applied to
realistic problems by giving a specific example.

\section{An application: wave packets in a wave guide}
\indent
As an application let us consider the propagation of an 
electromagnetic wave packet in a hollow wave guide, placed along 
the z axis, of arbitrary (but constant) cross-sectional shape with 
boundary surfaces being perfect conductors (the development of this 
paragraph is a generalization of chapter 8 of \cite{Jackson}). If 
$\omega$ is the frequency of the incident signal, let us write the 
spatial and temporal dependence of the electric and magnetic field 
inside the guide as
\bea
\bvec{E} (x,y,z,t) & = & \bvec{E_0} (x,y,\zeta) \, e^{i (kz - \omega t)} 
\label{31}  \\
\bvec{B} (x,y,z,t) & = & \bvec{B_0} (x,y,\zeta) \, e^{i (kz - \omega t)} 
\label{32} 
\eea
where $k$ is an as yet unknown wave number and $\zeta$ is given in the 
previous section. The fields $\bvec{E_0}$,$\bvec{B_0}$ satisfy the wave 
equation
\be
\left( \nabla^2_{\perp} \, + \, \left( 
\frac{\omega^2}{c^2} \, - \, k^2 \right)
\, + \, 2i \left( k \, - \, \frac{v \omega}{c^2} \right) 
\frac{\partial}{\partial \zeta} \, + \, \left( 1 \, - \, 
\frac{v^2}{c^2} \right) \frac{\partial^2}{\partial \zeta^2} \right) \, 
\left( 
\ba{c}
\bvec{E_0} \\
\bvec{B_0}
\ea  \right) \; = \; 0
\label{33}
\ee
where $\nabla^2_{\perp} \, = \, \frac{\partial^2}{\partial x^2} \, + \, 
\frac{\partial^2}{\partial y^2}$. 

Now we impose the $\zeta$-dependence of the fields to be given by one 
of the quantification rules eq. (\ref{25}) or (\ref{28}); 
then the $x,y$ 
dependence is determined by the equation
\be
\left( \nabla^2_{\perp} \, + \, \frac{\omega^2}{c^2} \, - \, k^2
\, \mp \, \Omega^2 \right) \,
\left( 
\ba{c}
\bvec{E_0} \\
\bvec{B_0}
\ea  \right) \; = \; 0
\label{34}
\ee
where the upper sign refers to propagation with $v<c$ and the other one 
to $v>c$. From this follows that, considering for example 
subluminal group velocities, in terms of the transverse and parallel 
(respect to the z guide axis) components, Maxwell equations become
\footnote{In general what follows remains valid if the substitution 
$\Omega \rightarrow - \Omega$ is performed. For clearness of notation 
we restrict only to one sign.}
\be
\left( k \, + \, \gamma \Omega \right) \, \bvec{E_{\perp}} \, + \, 
\frac{1}{c} \, \left( \omega \, + \, \gamma \Omega v \right) \, 
\bvec{e_3} \times \bvec{B_{\perp}} \; = \; -i \, \bvec{\nabla_{\perp}} \, 
E_z
\label{35}
\ee
\be
\left( k \, + \, \gamma \Omega \right) \, \bvec{B_{\perp}} \, - \, 
\frac{1}{c} \, \left( \omega \, + \, \gamma \Omega v \right) \, 
\bvec{e_3} \times \bvec{E_{\perp}} \; = \; -i \, \bvec{\nabla_{\perp}} \, 
B_z
\label{36}
\ee
\be
\bvec{e_3} \cdot \left( \bvec{\nabla_{\perp}} \times \bvec{E_{\perp}} 
\right) \; = \; \frac{i}{c} \, \left( \omega \, + \, \gamma \Omega 
v \right) \, B_z  \;\;\;\;
\label{37}
\ee
\be
\bvec{e_3} \cdot \left( \bvec{\nabla_{\perp}} \times \bvec{B_{\perp}} 
\right) \; = \; - \, \frac{i}{c} \, \left( \omega \, + \, \gamma \Omega 
v \right) \, E_z
\label{38}
\ee
\be
\bvec{\nabla_{\perp}} \cdot \bvec{E_{\perp}} \; = \; - \, i
\left( k \, + \, \gamma \Omega \right) \, E_z
\label{39}
\ee
\be
\bvec{\nabla_{\perp}} \cdot \bvec{B_{\perp}} \; = \; - \, i
\left( k \, + \, \gamma \Omega \right) \, B_z
\label{310}
\ee
$\bvec{e_3}$ being a unit vector in the z direction (for $v > c$ it 
suffices to replace $\gamma \, = \, (1 - v^2/c^2)^{-1/2}$ with
$\tilde{\gamma} \, = \, (v^2/c^2 - 1)^{-1/2}$). We deduce that if 
$E_z$,$B_z$ are known, then from the first two equations 
(\ref{35}),(\ref{36}) the transverse components of $\bvec{E}$,$\bvec{B}$
are determined, once the $\zeta$-dependence is known from the 
quantification rules; this leads, with 
respect to the normal case in which 
$v=c$, to an ``effective'' propagation frequency (and wave number) 
inside the guide given by $\omega' \, = \, \omega \, + \, \gamma 
\Omega v$ (or $\omega' \, = \, \omega \, + \, \tilde{\gamma} 
\Omega v$ for $v > c$), quantified just by the quantification rules.

In fact, for TM waves, specified by the conditions
\bea
B_z & = & 0 \;\;\;\;\;\;\;\;\;\;\;\;\;\;\; \mrm{everywhere}
\nonumber \\
& & \label{311} \\
E_z & = & 0 \;\;\;\;\;\;\;\;\;\;\;\;\;\;\; \mrm{on \; the \; guide 
\; surface}
\nonumber
\eea
we have
\be
\bvec{E_{\perp}} \; = \; i \, \frac{k \, + \, \gamma 
\Omega}{\Gamma^2} \, \bvec{\nabla_{\perp}} \, E_z
\label{312}
\ee
while for TE waves, specified by the conditions
\bea
E_z & = & 0 \;\;\;\;\;\;\;\;\;\;\;\;\;\;\; \mrm{everywhere}
\nonumber \\
& & \label{313} \\
\frac{\partial B_z}{\partial n} & = & 0 \;\;\;\;\;\;\;\;\;\;\;\;\;\;\; 
\mrm{on \; the \; guide \; surface}
\nonumber
\eea
the transverse components of the magnetic field is
\be
\bvec{B_{\perp}} \; = \; i \, \frac{k \, + \, \gamma 
\Omega}{\Gamma^2} \, \bvec{\nabla_{\perp}} \, B_z
\label{314}
\ee
In every case
\be
\bvec{B_{\perp}} \; = \; Y \, \bvec{e_3} \times \bvec{E_{\perp}}
\label{315}
\ee
with
\bea
Y & = & \frac{\omega \, + \, \gamma \Omega v}{c (k \, + \, \gamma 
\Omega )} \;\;\;\;\;\;\;\;\;\;\;\;\; \mrm{TM \; waves}
\nonumber \\
& & \label{316} \\
Y & = & \frac{c (k \, + \, \gamma \Omega )}
{\omega \, + \, \gamma \Omega v} \;\;\;\;\;\;\;\;\;\;\;\;\; \mrm{TE 
\; waves}
\nonumber
\eea
In eqs. (\ref{312}), (\ref{314})
\be
\Gamma^2 \; = \; \frac{\omega^2}{c^2} \, - \, k^2 \, \mp \, 
\Omega^2
\label{317}
\ee
are the eigenvalues of the wave equation (\ref{34}) with the boundary 
condition given respectively by eq. (\ref{311}),(\ref{313}). For example, 
for a rectangular wave guide with cross-section $a \cdot b$, the 
eigenvalues for TE waves are \cite{Jackson}:
\be
\Gamma^2_{m n} \; = \; \pi^2 \, \left( \frac{m^2}{a^2} \, + \, 
\frac{n^2}{b^2} \right)
\label{318}
\ee
with $m,n$ positive integers. 

The equation (\ref{317}) is the dispersion relation for $k$: for an 
incident wave with a given $\omega$, the propagating modes are those 
with the wave number
\be
c^2 k^2 \; = \; \omega^2 \, - \, c^2 \left( \Gamma^2 \, \pm \, 
\Omega^2 \right)
\label{319}
\ee
First, let us consider the case in which $v < c$; the propagation inside 
the guide effectively happens only for real values of $k$, i.e. for 
frequencies $\omega$ greater than the ``cutoff'' frequency
\be
\omega_c \; = \; c \, \sqrt{\Gamma^2 \, + \, \Omega^2 }
\label{320}
\ee
while for $\omega < \omega_c$ we are in the evanescent waves regime. 
Note that the effective cutoff frequency $\omega_c$ is greater than 
the usual value $\omega_{0c} \, = \, c \Gamma$ ($\Omega \, = \, 0$) 
and furthermore, for evanescent modes, the z-dependence does not show a 
pure exponential decay but is of the type $e^{i \gamma \Omega z} e^{- 
|k| z}$ (obviously we take $\Omega$ always real). 

Now, let us consider the most interesting case for which $v > c$; we can 
easily prove  that no cutoff arise. In fact, calculating the group 
velocity from eq. (\ref{319})
\be
v \; = \; c \, \sqrt{1 \, - \, \frac{\omega^2_{0c}}{\omega^2} \, + \, 
\frac{c^2 \Omega^2}{\omega^2}}
\label{321}
\ee
we find that it is effectively $v > c$ only if the condition
\be
c^2 \Omega^2 \; > \; \omega^2_{0c}
\label{322}
\ee
is fulfilled. This condition makes that in eq. (\ref{319}) the wave 
number is always real. Note that if $ c^2 \Omega^2 \, < \, 
\omega^2_{0c}$ we again have propagation with $v < c$, with the same 
features as discussed above, but now the effective cutoff frequency
$\omega_c \, = \, c \, \sqrt{\Gamma^2 \, - \, \Omega^2 }$ is lower 
than the usual value $\omega_{0c}$.

The obtained results bring to a very important consequence: a group 
velocity (and then a traversal time) can always be coherently defined 
for $v > c$ propagation. Notice that the fact that $\Omega$ is a 
property of the experimental setup (see at the end of the previous 
section) guarantees the univocity of the identification $v_{gr} = 
v(\omega_0)$ ($v$ depending on $\Omega$) and so ambiguities in the 
definition of the traversal time are really not present.

\section{Discussion on photonic tunneling}
\indent
Recently Martin and Landauer \cite{Landauer} have shown that the 
propagation of electromagnetic evanescent waves in a wave guide can be 
viewed as a photonic tunneling process through a barrier; this 
facilitates the experimental study of tunneling phenomena, due to the 
charge neutrality of photons with respect to other particles such as 
electrons. From the theoretical point of view, an open question is 
that of the barrier traversal time, because in the barrier region the 
momentum of the tunneling particle is imaginary so that no velocity 
can be defined. There are several approaches to the problem 
\cite{Rev} leading to different definitions of the traversal time. 
Instead, experimentally this time can be univocally measured, for 
example, from the coincidence of two photons, one travelling through 
the barrier and the other travelling in vacuum.

Enders and Nimtz \cite{Nimtz,diss} 
have studied photonic tunneling by means 
of microwave transmission through undersized wave guides operating 
below their cutoff (of the order of 6 $\div$ 9 GHz) and have obtained 
traversal times (for opaque barriers) from pulsed measurements in the 
time domain or, indirectly, in the frequency domain. Alternatively, 
Steinberg, Kwiat and Chiao \cite{Stein} and later Spielmann, Szip\"ocs 
and Krausz \cite{Spiel} employed some 1D photonic band-gap material as 
barrier for measuring tunneling times, in the UV and optical region 
respectively, with the aid of a two-photon interferometer.

The experimental evidence can be summarized as follows. Both in 
microwave and in photonic band-gap experiments photonic tunneling is 
observed; if the barrier medium is non-dissipative, the traversal time 
in this opaque region is nearly independent of the barrier thickness 
(Hartman effect \cite{Hartman}), so that for particular values of 
this length superluminal group velocities have been inferred. 
Furthermore, two strange properties have been detected. First, in 
microwave experiments, dissipative tunneling studies have shown that 
the Hartman effect disappears with increasing dissipation \cite{diss}. 
Second, 
the measurements of the tunneling of optical pulses through photonic 
band-gaps reveal that the pulses transmitted through a particular 
sample are significatively {\it shorter} than the incident ones
\cite{Spiel}; this 
effect disappears for increasing transmission coefficients. The latter 
two effects seem to indicate, in our opinion, that {\it real} 
propagation in the opaque region happens, and this stimulates to apply 
our theory to the present case. 
%However, we point out that the 
%formalism developed in section 2 and 3 is fully classical, so that for 
%applications to photonic tunneling some quantum effects should be
%considered.
%In any case, it is intriguing the fact that some results can 
%be applied to describe experimental evidence ``just so''. 
For example, 
we stress that a usual mass term (for $v < c$) or a tachionic mass 
\cite{Recami} for the photon (for $v > c$) 
cannot take into account the observed superluminal tunneling,
because of the dependence on the barrier thickness of 
the measured group velocity (i.e. the dependence on a characteristic 
parameter and not on an intrinsic one such as usual or tachionic 
mass). Instead, our quantification frequency $\Omega$ is not an 
intrinsic property, but would just depend on the employed experimental 
setup, so that the dependence on it of the group velocity (see for 
example eq. (\ref{321}) ) seems to go in the right direction for 
taking into account the superluminal tunneling. Moreover, the simple fact 
that in our approach a group velocity can always be defined for 
superluminal propagation eliminates the ambiguities in the definition 
of the traversal time. 

On the other hand, our theory can be tested independently from 
photonic tunneling, for example constructing ``ad hoc'' an 
electromagnetic apparatus which realizes 
eq. (\ref{25}) or (\ref{28}) in some regions and then measuring the 
dispersion relation between $\omega$ and $k$  or the group velocity of 
the propagating waves.

\section{Conclusions}
\indent
We have studied the propagation with group velocity $v_{gr} \neq 
c$ of the solutions of Maxwell equations (in vacuum) and have 
shown that it is possible (and not violating Einstein causality) 
provided a peculiar space-temporal dependence of the wave amplitude 
is given through a definite quantification rule.

As an application of the presented formalism we have considered the 
propagation in a wave guide and obtained, for $v_{gr} < c$, a different
effective cutoff frequency respect to the normal case, while for $v_{gr} > c$ 
no effective cutoff arises so that the wave number is always real and 
a group velocity can always be defined. In the evanescent regime for 
the $v_{gr} < c$ case, moreover, the real propagating waves show not a pure 
exponential decay but are only damped waves. 

Discussing photonic tunneling, we have pointed out how the presented 
approach can qualitatively describe the experimental evidences on this 
effect, even if definitive conclusions are not yet reached and further 
experimental and theoretical investigations are needed.

\vspace{1truecm}
\noindent
{\Large \bf Acknowledgements}\\
\noindent
The author is very grateful to Prof. F.Buccella and Prof. S.Solimeno
for helpful discussions and to Prof. E.Recami for his unfailing 
encouragement and very useful talks. Furthermore, the author is 
indebted with an anonymous referee for his very interesting comments 
and constructive criticism.


\begin{thebibliography}{99}

\bibitem{Levi}
T.Levi-Civita, ``Caratteristiche dei sistemi differenziali e 
propagazione ondosa'' (Zanichelli, Bologna, 1988). 

\bibitem{Bosanac}
S.Bosanac, {\it Phys. Rev.} {\bf A 28} (1983) 577. 

\bibitem{Batetal}
H.Bateman, ``Electrical and Optical Wave Motion'' (Cambridge 
University Press, Cambridge, 1915);\\
A.O.Barut and A.J.Brachen, {\it Found. Phys.} {\bf 22} (1992) 1267;\\
R.Donnelly and R.Ziolkowsky, {\it Proc. R. Soc. London} {\bf A 437} 
(1992) 673;\\
R.Donnelly and R.Ziolkowsky, {\it Proc. R. Soc. London} {\bf A 440} 
(1993) 541;\\
R.Ziolkowsky, I.M.Besieris and A.M.Sharavi, {\it J. Opt. Soc. Am.} 
{\bf A 10} (1993) 75.

\bibitem{RodVaz}
W.A.Rodrigues, jr. and J.Vaz, jr., Preprint {\bf hep-th/9511182}.

\bibitem{Nimtz}
A.Enders and G.Nimtz, {\it J. Phys.} I (France) {\bf 2} (1992) 1693;\\
A.Enders and G.Nimtz, {\it Phys. Rev} {\bf B 47} (1993) 9605;\\
A.Enders and G.Nimtz, {\it Phys. Rev} {\bf E 48} (1993) 632;\\
A.Enders and G.Nimtz, {\it J. Phys.} I (France) {\bf 3} (1993) 1089;\\
G.Nimtz, A.Enders and H.Spieker, {\it J. Phys.} I (France) {\bf 4} 
(1994) 565.

\bibitem{diss}
G.Nimtz, H.Spieker and H.M.Brodowsky, {\it J. Phys.} I (France) {\bf 4} 
(1994) 1379.

\bibitem{Stein}
A.M.Steinberg, P.G.Kwiat and R.Y.Chiao, {\it Phys. Rev. Lett.} {\bf 
71} (1993) 708.

\bibitem{Spiel}
Ch.Spielmann, R.Szip\"ocs, A.Stingl and F.Krausz, {\it Phys. Rev. Lett.} {\bf 
73} (1994) 2308.

\bibitem{Ranfagni}
A.Ranfagni, D.Mugnai, P.Fabeni and G.P.Pazzi, {\it Appl. Phys. Lett.} 
{\bf 58} (1991) 774;\\
A.Ranfagni, D.Mugnai, P.Fabeni and G.P.Pazzi, {\it  Phys. Rev.} 
{\bf E 48} (1993) 1453;\\
A.Ranfagni, D.Mugnai, P.Fabeni, G.P.Pazzi, G.Naletto and C.Sozzi, 
{\it Physica} (Amsterdam) {\bf B 175} (1991) 283.

\bibitem{Causality}
W.Heitmann and G.Nimtz, {\it Phys. Lett.} {\bf A 196} (1994) 154.

\bibitem{Brillou}
L.Brillouin, ``Wave propagation and group velocity'' (Academic Press, 
New York, 1960).

\bibitem{Rev}
V.S.Olkhovsky and E.Recami, {\it Physics Reports} {\bf 214} (1992) 339;\\
V.S.Olkhovsky. E.Recami, F.Raciti and A.K.Zaichenko, 
{\it J. Phys.} I (France) {\bf 5} (1995) 351.

\bibitem{Jackson}
J.D.Jackson, ``Classical Electrodynamics'' (Wiley, New York, 1975).

\bibitem{Landauer}
Th.Martin and R.Landauer, {\it Phys. Rev.} {\bf A 45} (1991) 2611.

\bibitem{Hartman}
Th.E.Hartman, {\it J. Appl. Phys.} {\bf 33} (1962) 3427.

\bibitem{Recami}
E.Recami, {\it Riv. Nuovo Cimento} {\bf 9} (1986), issue no. 6, pp. 1-178, 
and references therein.

\end{thebibliography}
\end{document}